%% file: GapJunctions.tex
\documentclass[final,copyright,creativecommons]{eptcs}
\providecommand{\event}{HSB 2012}

\usepackage{cite}
\usepackage{graphicx}
\usepackage{parskip}
\usepackage{amsmath}
\usepackage{amssymb}
\usepackage[framed,numbered]{mcode}
\usepackage{subfig}
\usepackage[compact]{titlesec}

\titlespacing{\section}{0pt}{*0.25}{*0.25}
\titlespacing{\subsection}{0pt}{*0.25}{*0.25}
\titlespacing{\subsubsection}{0pt}{*0.25}{*0.25}

\title{Modelling the effect of gap junctions on tissue-level cardiac electrophysiology}
\author{{\centering \quad \quad Doug Bruce  \qquad \qquad \qquad Pras Pathmanathan \ \ \qquad \qquad Jonathan P.\ Whiteley}
\email{{\centering douglas.bruce@oriel.ox.ac.uk \quad pras.pathmanathan@cs.ox.ac.uk \quad jonathan.whiteley@cs.ox.ac.uk}}
\institute{Computational Biology Group, Department of Computer Science, University of Oxford}
}

\def\titlerunning{Modelling the effect of gap junctions on tissue-level cardiac electrophysiology}
\def\authorrunning{D.A.\ Bruce,  P.\ Pathmanathan \& J.P.\ Whiteley}
\begin{document}
\maketitle

\abovedisplayshortskip=-14pt
\belowdisplayshortskip=-2pt
\abovedisplayskip=-2pt
\belowdisplayskip=10pt

\vspace{-10pt}

\begin{abstract}

When modelling tissue-level cardiac electrophysiology, continuum approximations to the discrete cell-level equations are used to maintain computational tractability. One of the most commonly used models is represented by the bidomain equations, the derivation of which relies on a homogenisation technique to construct a suitable approximation to the discrete model. This derivation does not explicitly account for the presence of gap junctions connecting one cell to another. It has been seen experimentally [Rohr, Cardiovasc.\ Res.\ 2004] that these gap junctions have a marked effect on the propagation of the action potential, specifically as the upstroke of the wave passes through the gap junction.

In this paper we explicitly include gap junctions in a both a 2D discrete model of cardiac electrophysiology, and the corresponding continuum model, on a simplified cell geometry. Using these models we compare the results of simulations using both continuum and discrete systems. We see that the form of the action potential as it passes through gap junctions cannot be replicated using a continuum model, and that the underlying propagation speed of the action potential ceases to match up between models when  gap junctions are introduced. In addition, the results of the discrete simulations match the characteristics of those shown in Rohr 2004. From this, we suggest that a hybrid model --- a discrete system following the upstroke of the action potential, and a continuum system elsewhere --- may give a more accurate description of cardiac electrophysiology.

\end{abstract}


\setlength{\parskip}{6pt}

\section{Introduction}

Many phenomena in biology are discrete; for example, biological tissue consists of discrete cells within extracellular material. When modelling a particular phenomenon or feature, different sets of equations can apply in the intra- and extracellular regions. In the case of modelling at the tissue or organ level it is impractical to model each individual cell. As a result, multi-scale techniques which consider the average behaviour of the problem --- for example homogenisation --- are used so that we may include cell-level phenomena into a tissue-level model whilst retaining computational tractability.

In the case of cardiac electrophysiology, whilst computing a numerical approximation to the solution of the governing equations at the level of individual cells should give accurate and realistic results, it is computationally unfeasible even for very small regions of tissue. It is much more efficient to solve these models using a continuum approximation to the equations, and indeed over the last 20 years many have done this \cite{chaste_2010}. This continuum model relies on a homogenisation technique to construct a suitable approximation to the discrete model. The homogenisation process uses a multiple-scales method which uses the fact that the problem is naturally defined on two different scales.

We now discuss relevant issues from cardiac electrophysiology and homogenisation.

\newpage

\subsection{Cardiac tissue modelling}

\begin{figure}[htbp]
\begin{center}

\includegraphics[width = 0.5\linewidth,height =
0.325\linewidth]{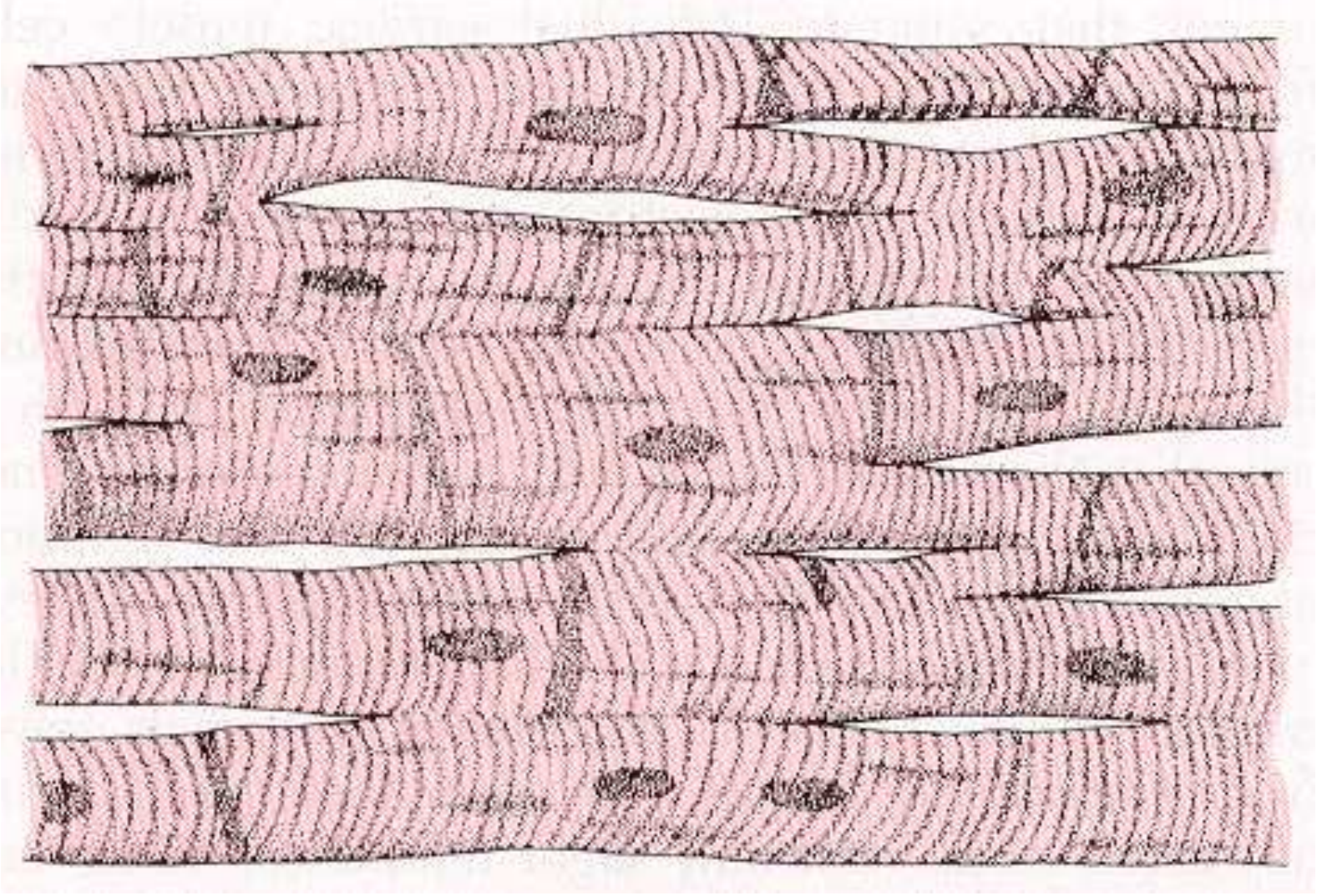}

\caption{\emph{Cardiac cell structure: Guyton and Hall, 1996, Fig.\ 9-2,
p.108.}}
\label{cardiaccells}
\end{center}
\end{figure}

An example of the histology of cardiac cells is shown in Figure \ref{cardiaccells}. These myocardial cells are roughly cylindrical and are packed together in an irregular three-dimensional pattern. Each cell is electrically connected to its neighbours via gap junctions, which are small channels through which ions may flow. The cell structure is surrounded by an extracellular matrix (ECM) and the two regions are separated by the cell membrane. In the discrete formulation of the governing equations, as will be described in more detail shortly, Laplace's equation holds for the potentials in both the intracellular and extracellular spaces, with the electrical properties of the membrane taken into account in the boundary conditions. The solution of these equations is a substantial computational task even for a small region of tissue due to the fine-scale structural detail of cells. As a result, it is computationally desirable to homogenise the microstructure to allow us to pose the problem as a continuum rather than a discrete problem.  This led to the proposal of the bidomain model by Tung in 1978 \cite{tung_1978} in which the homogenised potentials are solved for via two reaction-diffusion equations using averaged conductivities, where the precise form of averaging arises from the homogenisation technique. A sketch of the derivation of the bidomain equations is presented later in this document (see Section \ref{continuum}), and as discussed earlier it is the generally accepted model of cardiac tissue behaviour. A more formal derivation of the model can be found in \cite{keener_panfilov_1996}.

Whilst realistic simulations using the bidomain model have been demonstrated, these studies tend to take the bidomain equations as being physiologically correct in all circumstances, and empirical parameters are measured by matching the results of the continuum simulations with experimental data. However, there has been little rigorous testing of the validity of the derivation of the bidomain equations; in particular, concerning the homogenisation technique used to average microstructural quantities. 

In addition, the effect of gap junctions on propagation is often ignored during the homogenisation process. As previously mentioned, gap junctions are the means by which cells are electrically connected to one another. Whilst they allow the signal to be conducted, they do so with more resistance than is given by the interior of the cells. When considering action potential propagation at cell-level, we therefore expect to see the conduction velocity reduce as the wave passes through the gap junction. Indeed, this is supported by experimental data as shown in \cite{rohr_2004} --- the spatial form of the action potential is `stepped', with the reduced conductivity in the gap junction causing a steep jump in membrane potential between one side of the gap junction and the other. The repercussions of this for the continuum model, and how it affects the homogenisation process, will be discussed later.

\subsubsection{Hybrid models of cardiac electrophysiology}

In recent years, models based on hybrid automata have been developed to model networks of excitable cells, with an initial focus on the temporal morphology of the action potential \cite{ye_grosu_et_al_2005}. The model was further refined to study more specific conditions associated with action potential morphology such as early depolarisation \cite{ye_grosu_et_al_2008}, spiral waves \cite{grosu_et_al_2008}, and tachycardia \cite{grosu_et_al_2011}. Whilst these models allow efficient and precise analysis of the conditions required for such phenomena, they are not focused on the incorporation of the cell microstructure and the corresponding effect on action potential propagation.

\subsection{Homogenisation}

Whilst the initial formulation of the bidomain model stated the equations in the same form as they remain today (as laid out in \cite{tung_1978}), no rigorous derivation was given at that time that followed from basic physical principles and the cell-level properties of the tissue --- the macroscopic bidomain equations were not directly connected to the microstructure of the tissue. The advent of more formal homogenisation methods \cite{bensoussan_1979} paved the way for a first attempt at a rigorous derivation of the bidomain equations by Neu \& Krassowska in 1993 \cite{neu_krassowska_1993}. They used a multiple-scales asymptotic expansion technique to convert the microscopic problem, formulated in terms of the pointwise potentials, to a macroscopic problem formulated in terms of the leading order potential averages.

This technique was then modified by Keener \& Panfilov in 1996 \cite{keener_panfilov_1996}, who corrected a couple of errors in the homogenisation technique used. This boiled down to generalising the work of Neu \& Krassowska using more realistic tissue geometries. The error corrections played no part during normal action potential situations, but had a significant effects when examining the result of the application of a large current stimulus such as is seen during defibrillatory shocks, with the result being that the mechanism for defibrillation was clarified. It is a version of this derivation that will be presented later in this document.

Following this, a recent paper by Richardson \& Chapman \cite{richardson_chapman_2011} augmented the above derivation by introducing a co-ordinate transformation that allows for variable tissue structure and ultimately a set of bidomain equations in which the conductivity tensors systematically account for deformation of the tissue and the orientation of the cells.

\subsection{Analysis of the homogenisation process}

Despite all this work having been done on the derivation of the bidomain equations, looking at multiple-scales methods and complex geometry, one unproven assumption remains. In the derivations, the macro- and microscales of the problem are related by a dimensionless parameter, usually denoted $\epsilon$, that is equal to the ratio of the typical lengthscale of a single cell to the lengthscale over which the solution varies. In fact, the denominator of $\epsilon$ is stated as the `natural' lengthscale of the fibres in \cite{keener_panfilov_1996}, which has been translated in \cite{richardson_chapman_2011} as the `typical lengthscale of the cardiac tissue', but as we have stated above it is more correct to translate it as the typical solution lengthscale, so that $\epsilon$ is equal to the ratio between microscale and macroscale coordinates. This parameter is assumed to be small, and much of the homogenisation process involves taking the limit $\epsilon \to 0$. 

However, it is clear that in the case of a steep propagating wavefront, this will not be the case. As previously discussed, when the action potential passes through a gap junction the steepness of the upstroke portion of the wave will be significantly increased. We therefore question whether, in such circumstances, the homogenisation process used in the derivation of the bidomain equations remains valid.

Furthermore, it will become apparent when presented with an explicit version of the homogenisation process in Section \ref{continuum} that the `stepped' action potential seen in the presence of gap junctions cannot be captured using such a continuum model --- not only does the homogenisation process enforce that the macroscale conductivity tensors are independent of space at a cellular level, but the fundamental principle of the continuum model is that the fine-scale structure of cells, including gap junctions, are only accounted for by their average effect on the system.

\subsection{Aims and Outline}

We have described how the presence of gap junctions in cardiac tissue has a major effect on the propagation of the action potential and the associated conduction velocity. We therefore wish to consider the most accurate method to incorporate such structures into a simplified discrete model of cardiac electrophysiology. 

We begin in Section \ref{CardiacModelling} by describing the conventional discrete and continuum models of cardiac electrophysiology. Then, in Section \ref{GapJcts}, we consider the physiological properties of gap junctions and relate these to the models, setting up a more detailed discrete model that accounts for gap junctions, as well as deriving the corresponding continuum equations, which we will refer to as the modified bidomain equations. We then compare the results of these continuum and discrete models in Section \ref{Results}, asking how well the solutions compare both to each other and to those seen experimentally. We will observe that whilst the bidomain equations adequately represent the behaviour of a discrete model that neglects gap junctions, when gap junctions are included the continuum model wavespeed begins to fail to match the discrete model wavespeed, and that the discrete model better captures action potential profiles observed experimentally.


\section{Modelling cardiac electrophysiology} \label{CardiacModelling}

The derivation of the models below applies to general periodic structures. However, in this paper we are restricting our considerations to 2D models, and so we will now present a simplified representation of the cell structure in two dimensions. This will be the geometry upon which our analysis and simulations will take place.

\begin{figure}[htbp]
\begin{center}
\subfloat[A representation of cardiac cells in 2D. $\sigma_i$ and $\sigma_e$
are, respectively, the intracellular and extracellular conductivities, with
$\phi_i$ and $\phi_e$ the potentials. The region $\Omega$ represents a periodic
subunit containing both intracellular and extracellular
space.]{\label{RectangularCells}\input{fulldomain.latex}}
\qquad
\subfloat[A single periodic subunit of cell and ECM, as represented by $\Omega$
in Figure \ref{RectangularCells}. \textrm{L} represents the length of our subunit,
with $h$ its height and $h_1$ the height of the intracellular portion. The
intracellular domain is labelled $\Omega_i$, and the extracellular domain
$\Omega_e$. The membrane between the two is labelled $\partial
\Omega_m$.]{\label{Subunit}\input{subunit.latex}}
\caption{\emph{The 2D representation of cardiac cells}}
\label{2DCells}
\end{center}
\end{figure}
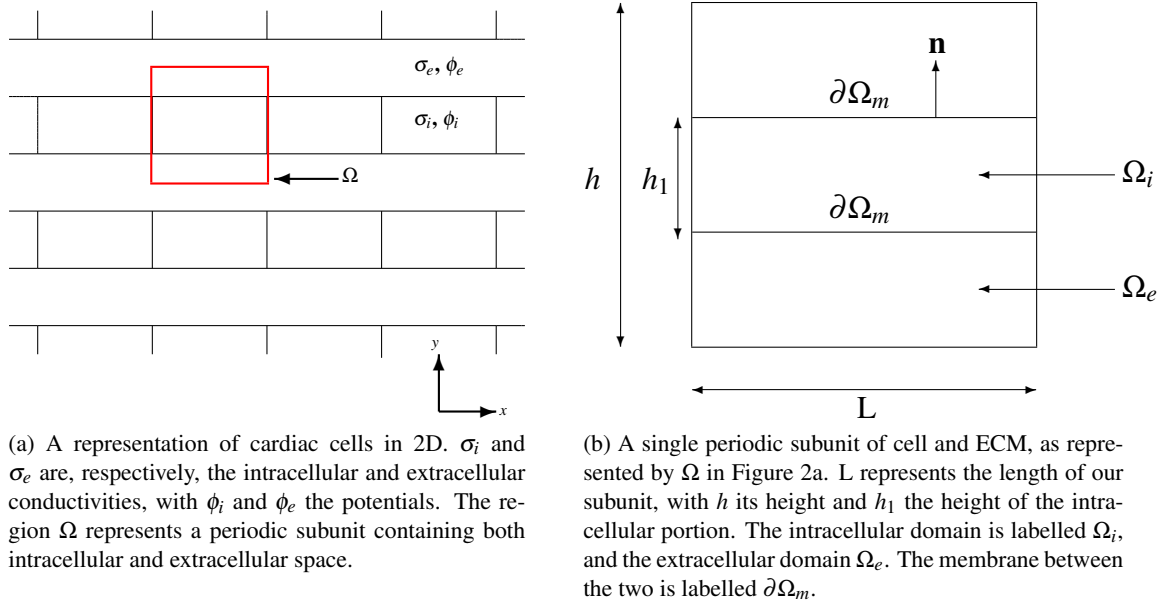

To arrive at our simplified representation of the cells, we assume that the intracellular space in each periodic subunit is rectangular in shape, with no coupling of the cells in the off-fibre direction, thus restricting our attention to propagation in one spatial dimension. This leads us to the schematic representation of Figure~\ref{RectangularCells}.

This representation consists of a sheet of cells that are connected into fibres along the $x$-direction, with each fibre separated in the $y$-direction by extracellular matrix. We create a periodic subunit of the domain, labelled $\Omega$, which contains both intracellular and extracellular portions, and this can be seen in more detail in Figure~\ref{Subunit}. Here, we are assuming that all cells have the same dimensions. 

\subsection{The discrete model} \label{discrete}

In this section, we are referring to the global problem as given in Figure \ref{RectangularCells}.

\subsubsection{Intracellular space}

In the intracellular space $\Omega_i$, the current is given by $\textbf{i}_c = -\sigma_i \nabla \phi_i$, where $\phi_i$ is the intracellular potential and $\sigma_i$ is the (scalar) conductivity of the intracellular space. The form of the current comes from Ohm's law which states that current is the product of conductivity and the gradient of the potential \cite{keener_sneyd_2001}. Conservation of current in the intracellular space therefore gives us

\begin{equation} \label{current_intra}
\nabla \cdot (\sigma_i \nabla \phi_i) = 0, \qquad \textbf{x} \in \Omega_i.
\end{equation}

The boundary condition representing flux of current is then

\begin{equation} \label{boundary_intra}
- \sigma_i \nabla \phi_i \cdot \textbf{n} = I_m(\textbf{x}), \qquad \textbf{x}
\in \partial \Omega_m,
\end{equation}

where $\textbf{n}$ is the outward pointing normal, that is, the normal pointing from the intracellular space into the extracellular space, $I_m$ is the transmembrane current, \emph{i.e.}\ the current flowing from the intracellular space into the extracellular space, and $\partial \Omega_m$ is the boundary between the intracellular and extracellular spaces.

\subsubsection{Extracellular space}

Similarly, in the extracellular space $\Omega_e$ we have

\begin{equation} \label{current_extra}
\nabla \cdot (\sigma_e \nabla \phi_e) = 0, \qquad \textbf{x} \in \Omega_e,
\end{equation}

where $\phi_e$ and $\sigma_e$ are the extracellular potential and conductivity respectively. The transmembrane current $I_m$ will now flow into the extracellular space, and so the boundary condition here becomes

\begin{equation} \label{boundary_extra}
\sigma_e \nabla \phi_e \cdot \textbf{n} = I_m(\textbf{x}), \qquad \textbf{x} \in
\partial \Omega_m,
\end{equation}

where $\partial \Omega_m$ is the same boundary and $\textbf{n}$ the same normal
as in \eqref{boundary_intra}, so that the normal still points from the intracellular
to the extracellular space.

\subsubsection{The transmembrane current}

For the transmembrane potential, defined on $\partial \Omega_m$ by $v_m = \phi_i - \phi_e$, the transmembrane current is given by

\begin{equation} \label{transmembrane_current}
I_m = C_m \frac{\partial v_m}{\partial t} + I_{ion}(v_m,\textbf{u}),
\end{equation}

where $C_m$ denotes the membrane capacitance and $\textbf{u}(\textbf{x},t)$ consists of various ionic concentrations and gating variables, which are determined from a cell model usually represented by a system of ordinary differential equations (ODEs). This comes from modelling the cell membrane as a capacitor, and as the capacitance of an insulator is defined to be the ratio between charge and potential we have $C_m =Q/v_m$.

Since the current is given by the rate of change of charge, \emph{i.e.}\ $\mathrm{d}Q/\mathrm{d}t$, it follows that the capacitive current is equal to $C_m \frac{\partial v_m}{\partial t}$, assuming that $C_m$ is a constant property of the material. In addition to this capacitive current there will be an ionic current created by the flow of ions through the membrane, and this is denoted $I_{ion}$. The total transmembrane current will be the sum of the capacitive and ionic currents, and is thus given by \eqref{transmembrane_current}. 

As stated previously, the quantity $I_{ion}$ represents the sum of the currents formed by the flow of charged ions across the cell membrane. These currents are due to the membrane possessing pore-forming proteins, known as \emph{channels}, which allow the passage of specific ions, for example sodium (Na$^+$) and potassium (K$^+$), down their electrochemical gradient. The specifics of the form of the current caused by these channels can be found in \cite{keener_sneyd_2001}. It is assumed that $I_{ion}$ is defined in the same fashion at all points on the cell membrane, and that there are a sufficiently large number of ion channels along the cell membrane for us to model $I_{ion}$ as a continuous function.

To summarise, in a discrete framework we will be solving the equations
[\eqref{current_intra},\eqref{current_extra},\eqref{transmembrane_current}] for
the unknowns $\phi_i$ and $\phi_e$, subject to the boundary conditions
[\eqref{boundary_intra},\eqref{boundary_extra}], as well as specifying that
solutions and current are continuous between one cell and the next.


\subsection{The continuum approximation} \label{continuum}

To model the cardiac tissue as a continuum, we use a homogenisation technique
based on the assumption that the lengthscale of the solution to the governing
equations is much larger than the length of an individual cell. We therefore
define a ``fast'' variable 

$$\textbf{z} = \frac{1}{\epsilon} \textbf{x},$$

where

$$\epsilon = \frac{\textrm{length of a single cell}}{\textrm{lengthscale of the
solution}},$$

and it is thus assumed that $\epsilon \ll 1$. Considering the intracellular
space to begin with, we can now write $\phi_i$ as a function of both
$\textbf{x}$ and $\textbf{z}$ and seek a solution by expanding in powers of
$\epsilon$, so that

\begin{equation} \label{expansion_intra}
\phi_i(\textbf{x},\textbf{z}) = \Phi_i(\textbf{x}) + \epsilon
\phi_{i1}(\textbf{x},\textbf{z}) + \epsilon^2\phi_{i2}(\textbf{x},\textbf{z}) +
\dots,
\end{equation}

where $\phi_{i1}, \phi_{i2}, \dots,$ are periodic in $\textbf{z}$ with zero
mean. A full version of the derivation can be found in \cite{neu_krassowska_1993} or \cite{keener_panfilov_1996}, but to summarise, by substituting \eqref{expansion_intra} into the discrete governing equations and boundary conditions, equating powers of $\epsilon$ and using the periodicity of $\phi_{i1}$ and $\phi_{i2}$, the equation for the intracellular potential can be written

\begin{equation} \label{intra}
\nabla_{\textbf{x}} \cdot (\Sigma_i \nabla_{\textbf{x}} \Phi_i) = 
\frac{1}{V_\textrm{cell}} \int_{\partial \Omega_m} I_m \ \mathrm{d}
S_{\textbf{z}},
\end{equation}

where $V_{\textrm{cell}}$ is the volume of our periodic subunit and $\Sigma_i$ is the macroscale intracellular conductivity tensor of the problem, given by

\begin{equation} \label{Sigma_i}
\Sigma_i = \frac{1}{V_{\textrm{cell}}}\int_{\Omega_i} \sigma_i \left(I +
\frac{\partial \textbf{W}^i}{\partial \textbf{z}}\right) \mathrm{d}
V_{\textbf{z}}.
\end{equation}

The functions $W^i_j$ are periodic in $\textbf{z}$ with zero mean, and satisfy

\begin{equation} \label{W}
\nabla_{\textbf{z}} \cdot (\sigma_i \nabla_{\textbf{z}}W^i_j) =
-\frac{\partial \sigma_i}{\partial z_j}, \quad \textbf{x} \in \Omega_i, \qquad \qquad
\nabla_{\textbf{z}} W^i_j \cdot \textbf{n} = -n_j, \quad
\textbf{x} \in \partial \Omega_m, \qquad \qquad j = 1,2\nonumber
\end{equation}

The same method, applied this time to the extracellular space, will give us 

\begin{equation} \label{extra}
\nabla_{\textbf{x}} \cdot (\Sigma_e \nabla_{\textbf{x}} \Phi_e) =  -
\frac{1}{V_\textrm{cell}} \int_{\partial \Omega_m} I_m \ \mathrm{d}
S_{\textbf{z}},
\end{equation}

where analogously

\begin{equation} \label{Sigma_e}
\Sigma_e = \frac{1}{V_{\textrm{cell}}}\int_{\Omega_e} \sigma_e \left(I +
\frac{\partial \textbf{W}^e}{\partial \textbf{z}}\right) \mathrm{d}
V_{\textbf{z}},
\end{equation}

The functions $W^e_j$ now satisfy

\begin{equation}
\nabla_{\textbf{z}} \cdot (\sigma_e \nabla_{\textbf{z}}W^e_j) =
-\frac{\partial \sigma_e}{\partial z_j}, \quad \textbf{x} \in \Omega_e, \qquad \qquad
\nabla_{\textbf{z}} W^e_j \cdot \textbf{n} = n_j, \quad
\textbf{x} \in \partial \Omega_m, \qquad \qquad j = 1,2\nonumber
\end{equation}

and are periodic in $\textbf{z}$ with zero mean. Thus, the
tissue-level conductivity tensors $\Sigma_i$ and $\Sigma_e$ will depend on the
domain shapes $\Omega_i$ and $\Omega_e$, the
volumne of our periodic subunit $V_{\textrm{cell}}$, the micro-level conductivity scalars $\sigma_i$
and $\sigma_e$, along with any quantity that will change the functions $\textbf{W}_{(1,2)}^{(i,e)}$.

\subsubsection{The bidomain equations}

If we now write the transmembrane potential $v_m$ as a power expansion in
$\epsilon$, so that

$$v_m = V_m(\textbf{x}) + \epsilon v_{m1}(\textbf{x},\textbf{z}) + \epsilon^2
v_{m2}(\textbf{x},\textbf{z}) + \dots,$$

we may express \eqref{intra} as

\vspace{-12pt} \begin{align}
& \nabla_{\textbf{x}} \cdot (\Sigma_i \nabla_{\textbf{x}} \Phi_i) & = & \
\frac{1}{V_{cell}} \int_{\partial \Omega_m} C_m \frac{\partial v_m}{\partial t}
+ I_{ion}(v_m,t) \ \mathrm{d} S_{\textbf{z}}, \nonumber \\ 
& & = & \ \chi \left(C_m \frac{\partial v_m}{\partial t} + \frac{1}{S_m}
\int_{\partial \Omega_m} I_{ion}(V_m(\textbf{x}) + \epsilon
v_{m1}(\textbf{x},\textbf{z}) + \epsilon^2 v_{m2}(\textbf{x},\textbf{z}) +
\dots,t) \ \mathrm{d} S_{\textbf{z}} \right), \nonumber
\end{align}\vspace{-20pt} 

where $S_m$ is the membrane surface area and $\chi = S_m/V_{cell}$. If we make
the assumption that we may ignore the contribution from $\epsilon^1$ and higher
order terms, \emph{i.e.}\ that

$$ \frac{1}{S_m} \int_{\partial \Omega_m} I_{ion}(V_m(\textbf{x}) + \epsilon
v_{m1}(\textbf{x},\textbf{z}) + \epsilon^2 v_{m2}(\textbf{x},\textbf{z}) +
\dots,t) \ \mathrm{d} S_{\textbf{z}} = I_{ion}(V_m(\textbf{x}),t)$$

which also physiologically implies that we are taking there to be sufficiently
many ion channels in each cell to model $I_{ion}$ and a continuous function as mentioned previously, we can write

\begin{equation} \label{intra_final}
\nabla_{\textbf{x}} \cdot (\Sigma_i \nabla_{\textbf{x}} \Phi_i) = \chi \left(
C_m \frac{\partial V_m}{\partial t} + I_{ion}(V_m,t) \right),
\end{equation}

and similarly \eqref{extra} becomes

\begin{equation} \label{extra_final}
\nabla_{\textbf{x}} \cdot (\Sigma_e \nabla_{\textbf{x}} \Phi_e) = - \chi \left(
C_m \frac{\partial V_m}{\partial t} + I_{ion}(V_m,t) \right).
\end{equation}

We now use the fact that $\phi_i = V_m + \phi_e$ to eliminate $\phi_i$ from \eqref{intra_final}, and denote the
transmembrane potential $V_m$ simply as $V$ to avoid any future subscript
confusion, to write \eqref{intra_final} and \eqref{extra_final} in the more familiar form of 

\begin{subequations} \label{bidomain}
\vspace{-12pt} \begin{align} 
\chi C_m \frac{\partial V}{\partial t} & = \nabla_{\textbf{x}} \cdot (\Sigma_i
\nabla_{\textbf{x}}(V + \phi_e)) - \chi I_{ion}, \\
\nabla_{\textbf{x}} \cdot ((\Sigma_i + \Sigma_e) \nabla_{\textbf{x}}\phi_e +
\Sigma_i \nabla_{\textbf{x}} V) & =  0.
\end{align}\vspace{-20pt} 
\end{subequations}

Appropriate boundary conditions, imposing zero flux on the boundary of the entire domain, are

\begin{equation}
-\Sigma_i \nabla (V+\phi_e) \cdot \textbf{n} = 0,  \qquad \qquad \Sigma_e \nabla \phi_e \cdot \textbf{n} = 0. \nonumber
\end{equation}

These are what we call the \emph{bidomain equations}, in the absence of any external stimuli.


\section{Explicitly incorporating gap junctions into the models} \label{GapJcts}

\subsection{Physiology of gap junctions}

\begin{figure}[htbp]
\begin{center}

\includegraphics[width = 0.6\linewidth,height = 0.4\linewidth]{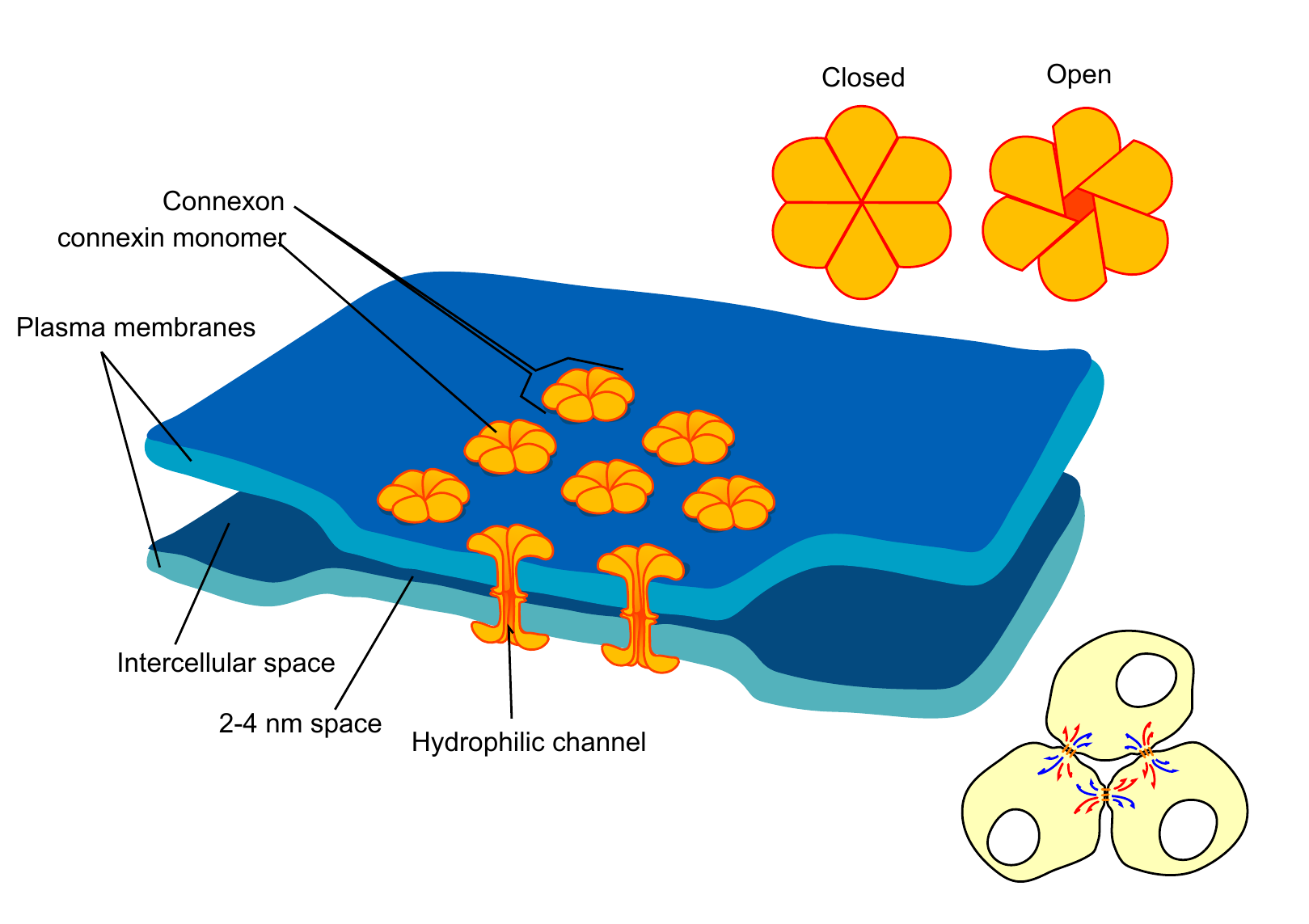}
\caption{\emph{A representation of gap junctions, Mariana Ruiz, 2006.}}
\label{GapJunctionWikipedia}
\end{center}
\end{figure}

\vspace{-1pt}

As seen in Figure \ref{GapJunctionWikipedia}, the gap junctions form a channel that directly connect two adjacent cells, allowing molecules and ions to pass through it. These junctions are abundant in cardiac muscle, and allow direct electrical signalling from one cell to the next. When looking at an entire cell, we can approximate the collection of individual gap junctions by one continuous domain whose conductivity $\sigma_g$ takes into account the density of gap junctions present, the average fraction that are open or closed during signal propagation and the underlying conductivity of the material.

We then notice that, on the boundary between such a gap junction domain and the intracellular spaces that it connects, both ions and electrical signals are free to flow, and so we will have continuity of flux across the boundary when considering electrical potential. This allows us to treat the gap junction as part of the intracellular space with a different conductivity, so that the intracellular conductivity becomes a function of space, which naturally corresponds to imposing continuity of potential and flux on the interface if the gap junction had been treated as a separate compartment.

On the interface between gap junction and extracellular space, we note that there are likely to be differences between the properties of this interface and that of the cell membrane. Unlike on the cell membrane, a gap junction will not allow ion transport between itself and the extracellular space. In addition, the capacitive properties of the material will be different from those of the cell membrane. These differences will have an impact on the form of the transmembrane current $I_m$ that will be defined on the interface between gap junction and extracellular space.

\newpage
\subsection{Adaptation of models to include gap junctions}
\vspace{-14pt}
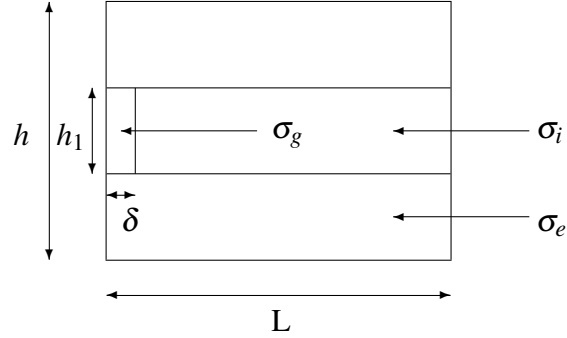
\begin{figure}[htbp]
\begin{center}
\input{gapjcts.latex}
\caption{\emph{Our periodic subunit (see Figure \ref{Subunit}) modified to include a
gap junction. It is modelled as a region of width $\delta$ at one end of the
cell with different conductivity ($\sigma_g$) to that of the cytoplasm ($\sigma_i$).}}
\label{GapJunctions}
\end{center}
\end{figure}

\vspace{-15pt}

From our discussion above, we adapt our cell geometry to include gap junctions as shown in Figure \ref{GapJunctions}. We include a thin region of width $\delta$ at one end of the intracellular space in which the conductivity is given by $\sigma_g$. The overall intracellular conductivity is now denoted by $\sigma = \sigma(\textbf{x})$, where

\begin{equation}
\sigma(\textbf{x}) = \begin{cases} \sigma_i & \textbf{x} \in \mathrm{cell}, \\
\sigma_g & \textbf{x} \in \mathrm{gap\ junction}. \end{cases}
\end{equation}

In addition, we adapt the form of the transmembrane current on the boundary as follows:

\begin{equation} \label{ModifiedTransmembraneCurrent}
I_m = \begin{cases} c_m  \frac{\partial v}{\partial t} + I_{ion} & \textbf{x} \in \mathrm{cell}, \\
c_g  \frac{\partial v}{\partial t} + I_g I_{ion} & \textbf{x} \in \mathrm{gap\ junction}. \end{cases}
\end{equation}

where $I_g$ is a boolean switch that turns ionic flow across the membrane on or off, and $c_g$ is the capacitance of the gap junction membrane. We may then alter the properties of the gap junction membrane by specifying $c_g$ according to one of three cases:

\begin{itemize}
\item $c_m$, treating it as if it were the same material as the cell membrane
\item $c_g \neq c_m$, treating it as a capacitive material with its own properties
\item 0, treating it as a fully insulative material.
\end{itemize}

Whilst we expect that it is correct to take $I_g=0$ and $c_g \neq c_m$ in the above system, we will leave both parameters undetermined in the forms stated so that we can explore the effect that they have on results of simulations.

To summarise, the discrete equations will be modified as such: the intracellular conductivity will now take different values in the cell and the gap junction, and the transmembrane current will have a different formulation on the cell membrane and the gap junction membrane.

\subsubsection{A modified formulation of the bidomain equations}

Returning to \eqref{intra}, the right-hand term of the equation given by $ \frac{1}{V_{cell}} \int_{\partial \Omega_m} I_m \mathrm{d} S $ will be split into integrals over the cell membrane $\partial \Omega_i$ and gap junction membrane $\partial \Omega_g$ separately before being evaluated, so

\vspace{-12pt} \begin{align}
\frac{1}{V_{cell}} \int_{\partial \Omega_m} I_m \mathrm{d} S & = \frac{1}{V_{cell}} \left[ \int_{\partial \Omega_i} I_m \mathrm{d} S + \int_{\partial \Omega_g} I_m \mathrm{d} S \right] \\
& =  \frac{1}{V_{cell}} \left[  c_i S_i \frac{\partial V}{\partial t} +  c_g S_g \frac{\partial V}{\partial t} +\int_{\partial \Omega_i} I_{ion} \mathrm{d} S + I_g\int_{\partial \Omega_g} I_{ion} \mathrm{d} S \right]\\ 
& =  (c_i \chi_i+c_g \chi_g)\frac{\partial V}{\partial t} + (\chi_i+ I_g\chi_g)I_{ion}
\end{align}\vspace{-20pt} 

where $c_g$ and $c_i$ are the capacitances of the gap junction and cell membranes, $S_g$ and $S_i$ are the surface areas of $\partial \Omega_g$ and  $\partial \Omega_i$  respectively, with $\chi_g = \frac{S_g}{V_{cell}}$ and $\chi_i = \frac{S_i}{V_{cell}}$. On our simplified geometry as given in Figure \ref{GapJunctions} we will have that $ \chi_i =\frac{2 (1-\delta )}{h}$ and $ \chi_g = \frac{2 \delta}{h}$.

The final form of the modified bidomain equations will therefore be

\begin{subequations} \label{modifiedbidomain}
\vspace{-12pt} \begin{align} 
(c_i \chi_i+c_g \chi_g) \frac{\partial V}{\partial t} & = \nabla_{\textbf{x}} \cdot (\Sigma_i
\nabla_{\textbf{x}}(V + \phi_e)) - (\chi_i+ I_g\chi_g) I_{ion}, \\
\nabla_{\textbf{x}} \cdot ((\Sigma_i + \Sigma_e) \nabla_{\textbf{x}}\phi_e +
\Sigma_i \nabla_{\textbf{x}} V) & =  0.
\end{align}\vspace{-20pt} 
\end{subequations}

\subsubsection{Effect on the intracellular conductivity tensor}

On a generalised geometry, we see that the tensor $\Sigma_i$ will be changed via the solutions to \eqref{W}, as the conductivity scalar there denoted $\sigma_i$ is modified to be a function of space.

On our simplified geometry, we are able to write down analytic solutions for $\Sigma_i$ in both cases. In the absence of gap junctions, the functions $W_1^i$ and $W_2^i$ given in \eqref{W} satisfy

$$\nabla^2 W^i_j = 0, \qquad j = 1,2.$$

The normal is given by $(0,1)$ and so the boundary conditions become

$$ \frac{\partial W_1^i}{\partial y} = 0, \qquad \mathrm{and} \qquad 
\frac{\partial W_2^i}{\partial y} = -1,$$

the solutions to which are

$$ W_1^i = A_1, \qquad W_2^i = -y + A_2,$$

where $A_1$ and $A_2$ and constants (note that they are not functions of $x$ as $W_1^i$ and $W_2^i$ are periodic in $x$). Substituting this into \eqref{Sigma_i} gives

\begin{equation} \label{Sigma_i_basiccell}
\Sigma_i = \frac{V_{\textrm{intra}}}{V_{\textrm{cell}}}\left(
\begin{array}{cc}
\sigma_i & 0\\
0 & 0
\end{array} \right),
\end{equation}

where $V_{\textrm{intra}}$ is the volume of the intracellular space.

In the presence of gap junctions, the equations for $W^i_2$ remain unchanged, and now $W^i_1$ will satisfy Laplace's equation in both the intracellular space and the gap junction separately with the same boundary condition as before. Therefore we have

\begin{equation}
\frac{\partial W^i_1}{\partial x} = \begin{cases} B_1 & 0 < x < \delta, \\
B_2  & \delta < x < \mathrm{L} . \end{cases}
\end{equation}

To find $B_1$ and $B_2$ we integrate the governing equation for $W^i_1$, given in \eqref{W}, across $x=\delta$ to give

$$\left[ \sigma \frac{\partial W^i_1}{\partial x} \right]^{\delta^+}_{\delta^-} = \left[ -\sigma \right]^{\delta^+}_{\delta^-},$$

and therefore that

$$ \sigma_i B_2 - \sigma_g B_1 = \sigma_g - \sigma_i.$$

We then use the fact that $W^i_1$ has zero mean in $x$ to give us

$$ \int_{\Omega_i} \frac{\partial W^i_1}{\partial x} \ \mathrm{d} x = \delta B_1 + (\mathrm{L}-\delta) B_2 = 0.$$

Solving for $B_1$ and $B_2$ and substituting into \eqref{Sigma_i} gives us that

$$ \Sigma_i^{(1,1)} = \frac{V_{\textrm{intra}}}{V_{\textrm{cell}}} \times \frac{\sigma_i \sigma_g \mathrm{L} }{\delta \sigma_i + (\mathrm{L}-\delta) \sigma_g}.$$

Note that, when $\sigma_g = \sigma_i$, this reduces to the corresponding entry in \eqref{Sigma_i_basiccell} as anticipated. It is also worth pointing out that, both with and without gap junctions included, we will have 

\begin{equation} \label{Sigma_e_basiccell}
\Sigma_e = \frac{V_{\textrm{extra}}}{V_{\textrm{cell}}}\left(
\begin{array}{cc}
\sigma_e & 0\\
0 & 0
\end{array} \right),
\end{equation}

where $V_{\textrm{extra}}$ is the volume of the extracellular portion of our periodic subunit.


\section{Results of simulations} \label{Results}

In order to investigate the effect of gap junctions on results of simulations, we solve both the continuum system, given by the modified bidomain equations \eqref{modifiedbidomain}, and the discrete system, given in [\eqref{current_intra},\eqref{boundary_intra},\eqref{current_extra},\eqref{boundary_extra}] with the transmembrane current given by \eqref{ModifiedTransmembraneCurrent}, using a Beeler-Reuter model for the ionic current \cite{beeler_reuter_1977} in both cases. We use a finite element method \cite{reddy_1993} with 320 nodes per cell in the discrete case and 80 nodes covering the corresponding area in the continuum case, and use the PETSc
library \\ (\texttt{http://www.mcs.anl.gov/petsc}) to solve the resulting linear systems.

We take an individual subunit to be of size 0.1 mm by 0.02 mm, with the intracellular portion 0.1 mm by 0.01 mm (so that, in the notation of Figure \ref{Subunit}, we have $L = 0.1$ mm, $h = 0.02$ mm and $h_1 = 0.01$ mm), using a 100 cell by 2 cell region. We simulate 50 ms of electrical activity, beginning the simulations in equilibrium so that $\phi_i = V_{eq}$ and $\phi_e = 0$  everywhere. Conduction coefficients are set at $\sigma_i$ = 0.175 $\mu$S/mm $\sigma_e$ = 0.7 $\mu$S/mm, and the membrane capacitance $c_m = 0.01 \ \mu$F/mm$^2$, with these parameters taken from \cite{chaste_2010}. We then apply an appropriate current stimulus, dependent on our solution parameters $\sigma_g$, $c_g$ and $I_g$, between 5ms and 10 ms to both cells on the $y$-axis.

The table below summarises the different parameter sets used in each of our simulations, along with a verbal characterisation of what we are simulating.

\begin{tabular}{c|c|c|c|c} 
\hline \hline
\multicolumn{5}{c}{\textbf{Parameter Values used in simulations}} \\ 
\hline
Model & $\sigma_g$ & $c_g$ & $I_g$ & Characterisation \\
\hline
Base & 0.175 & 0.01 & 1 & No gap junctions, models reduce to original forms \\
1 & 0.00175 & 0.01 & 1 & Gap junctions, simply reducing conductivity \\
2 & 0.00175 & 0.01 & 0 & Gap junctions, not allowing ion transport across membrane \\
3 & 0.00175 & 0.001 & 1 & Gap junctions, reduced capacitance  \\
4 & 0.00175 & 0.001 & 0 & Gap junctions, reduced capacitance, no ion transport \\
5 & 0.00175 & 0 & 1 & Gap junctions, fully insulating \\
6 & 0.00175 & 0 & 0 & Gap junctions, fully insulting, no ion transport \\
\hline \hline
\multicolumn{5}{c}{\label{Models}}
\end{tabular}

\subsection{Effect of introducing gap junctions}

\begin{figure}[htbp]
\begin{center}
\subfloat[]{\label{Base}\includegraphics[width = 0.45\linewidth,height =
0.3\linewidth]{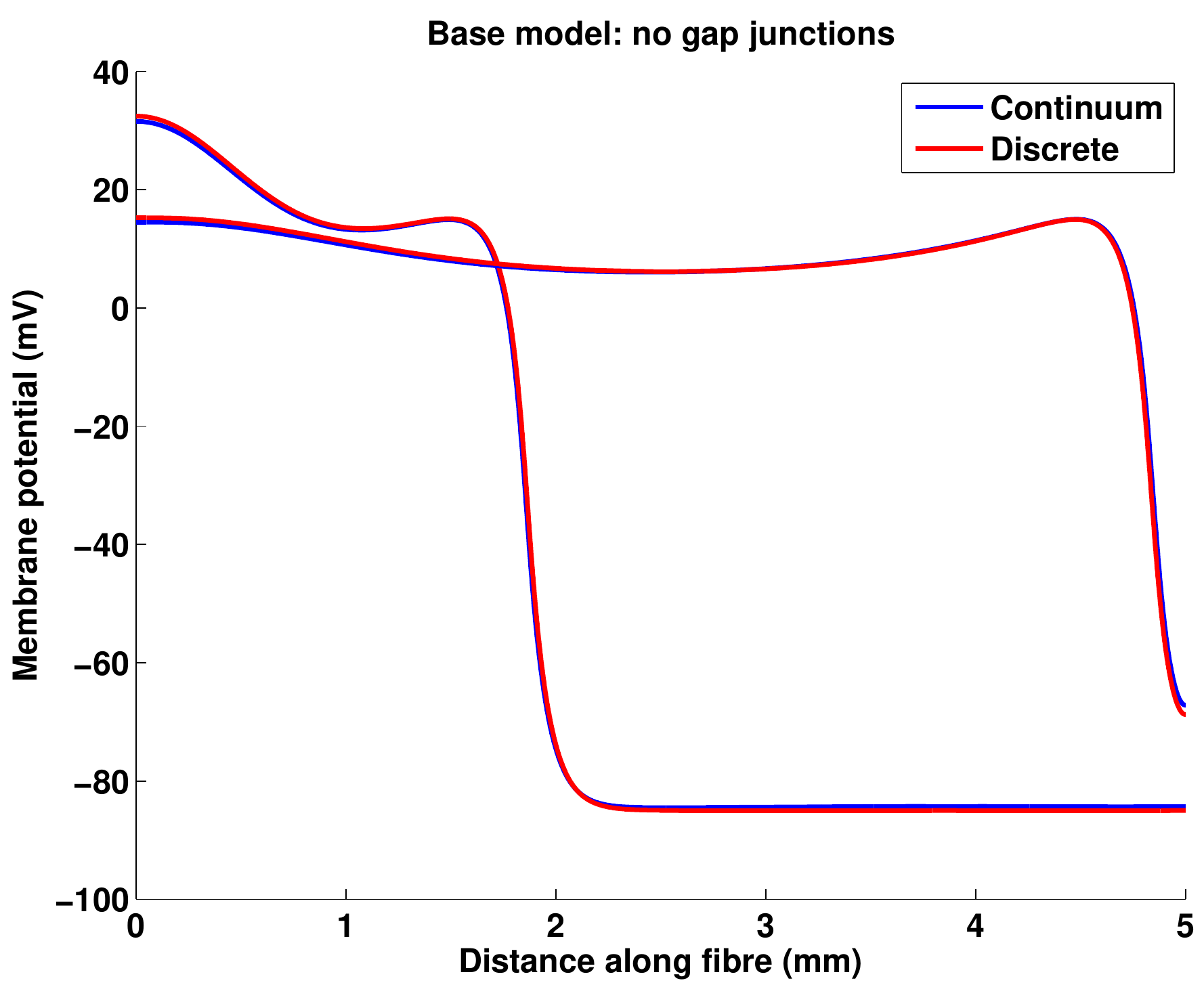}}
\subfloat[]{\label{Model1}\includegraphics[width = 0.45\linewidth,height =
0.3\linewidth]{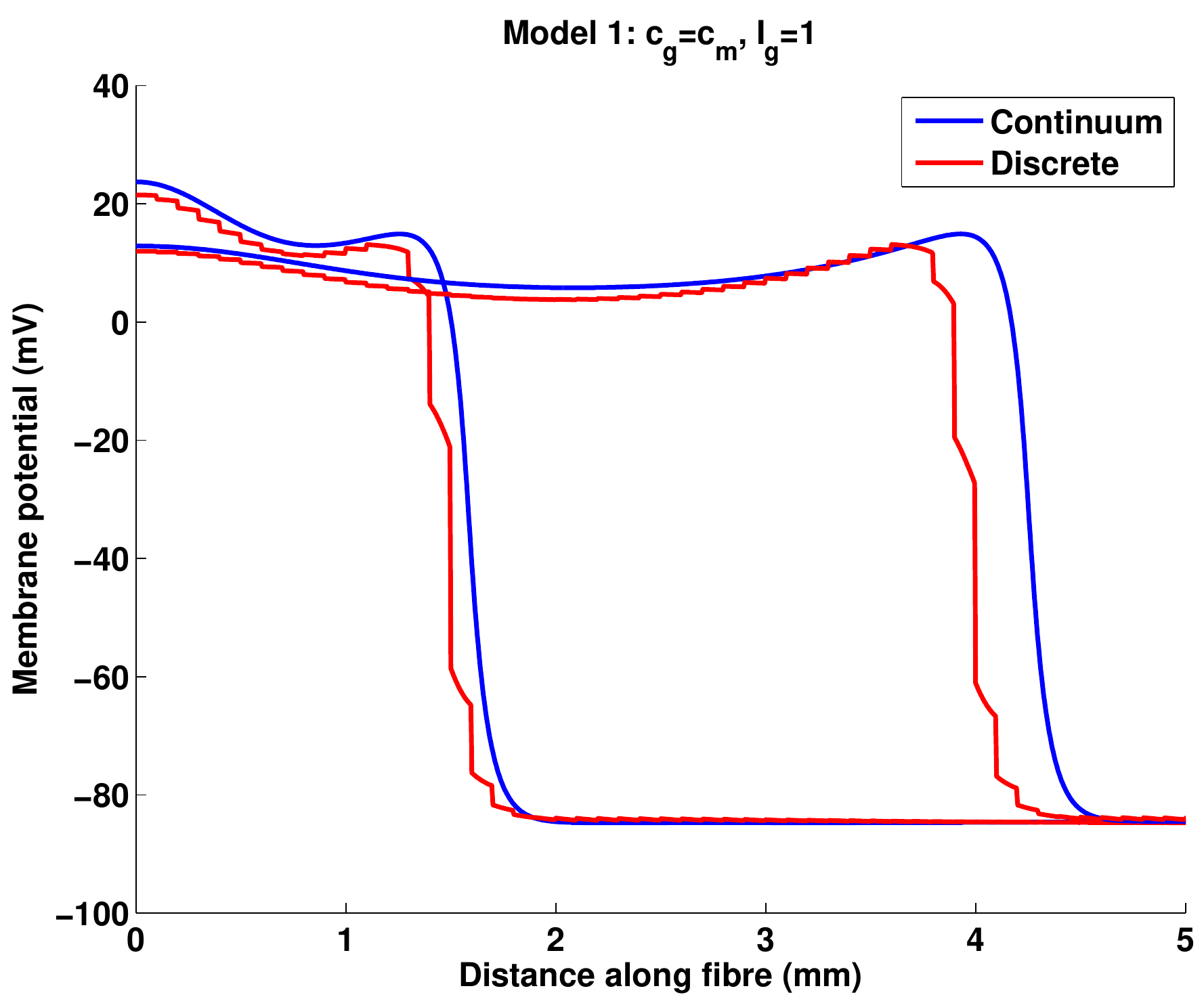}}
\caption{\emph{Results of simulations at 15 ms and 30 ms for our base case (left), in which gap junctions were not modelled, and our most simple implementation of gap junctions (right) in which they were treated as a region of reduced conductivity with identical properties to that of the cell.}}
\label{IntroduceGapJunctions}
\end{center}
\end{figure}

\begin{figure}[htbp]
\begin{center}
\subfloat[]{\label{ContinuumAll}\includegraphics[width = 0.45\linewidth,height =
0.3\linewidth]{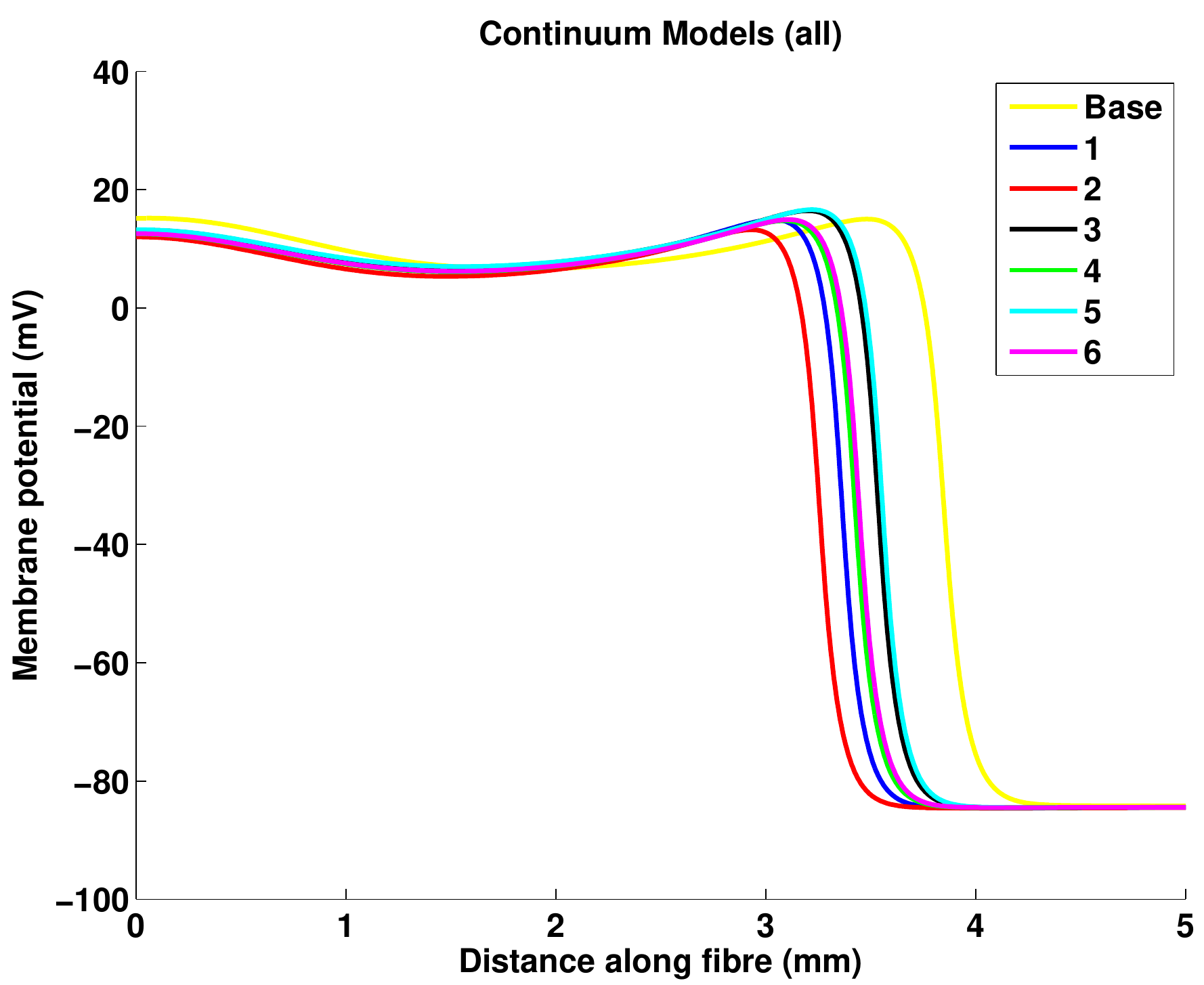}}
\subfloat[]{\label{DiscreteAll}\includegraphics[width = 0.45\linewidth,height =
0.3\linewidth]{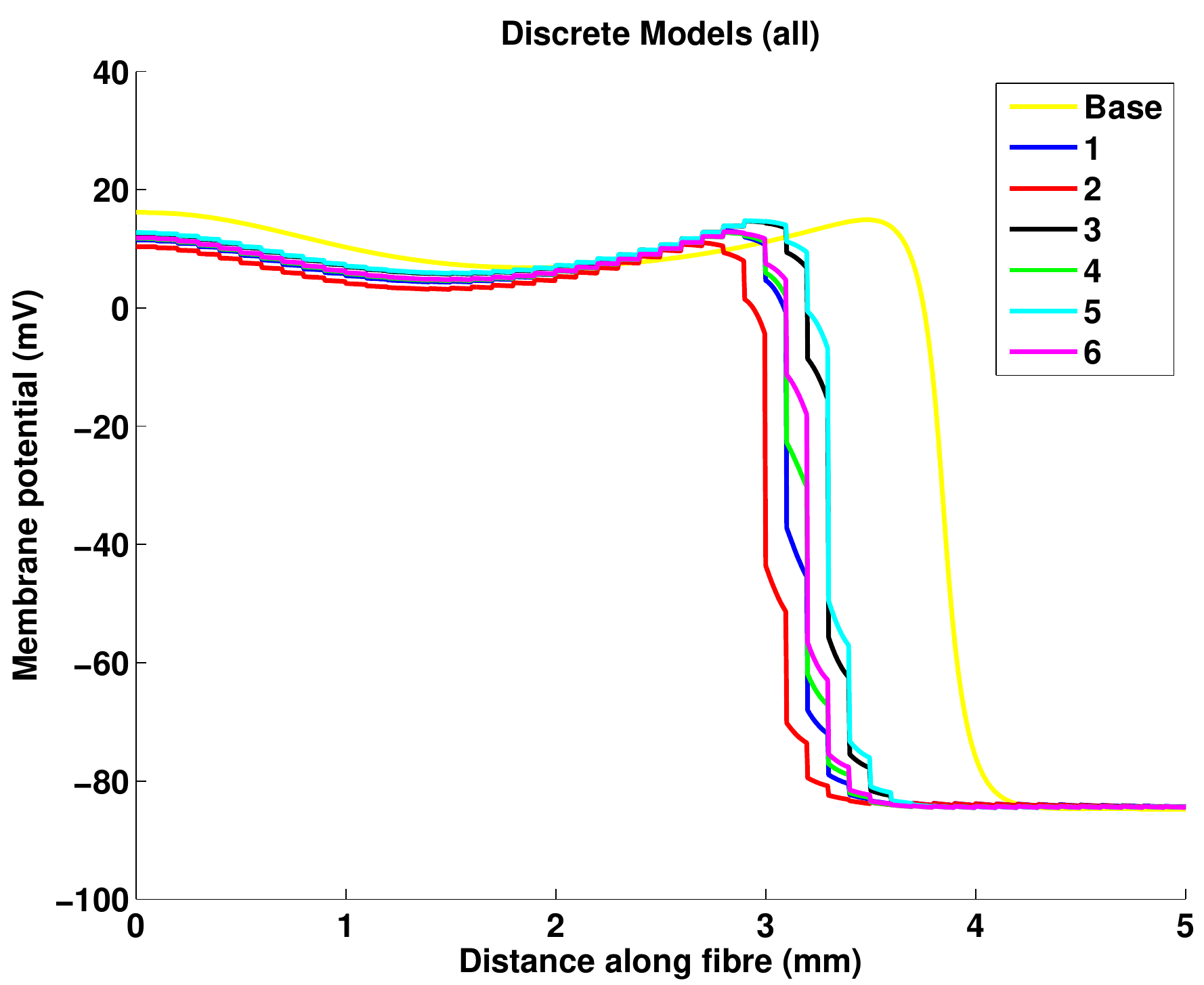}}
\caption{\emph{Results of simulations at 30 ms for all models, comparing the continuum solutions (left) and discrete solutions (right) against one another, to see how our choice of model affects propagation.}}
\label{CompareAll}
\end{center}
\end{figure}

\begin{figure}[htbp]
\begin{center}
\subfloat[]{\label{ContinuumZoom}\includegraphics[width = 0.45\linewidth,height =
0.3\linewidth]{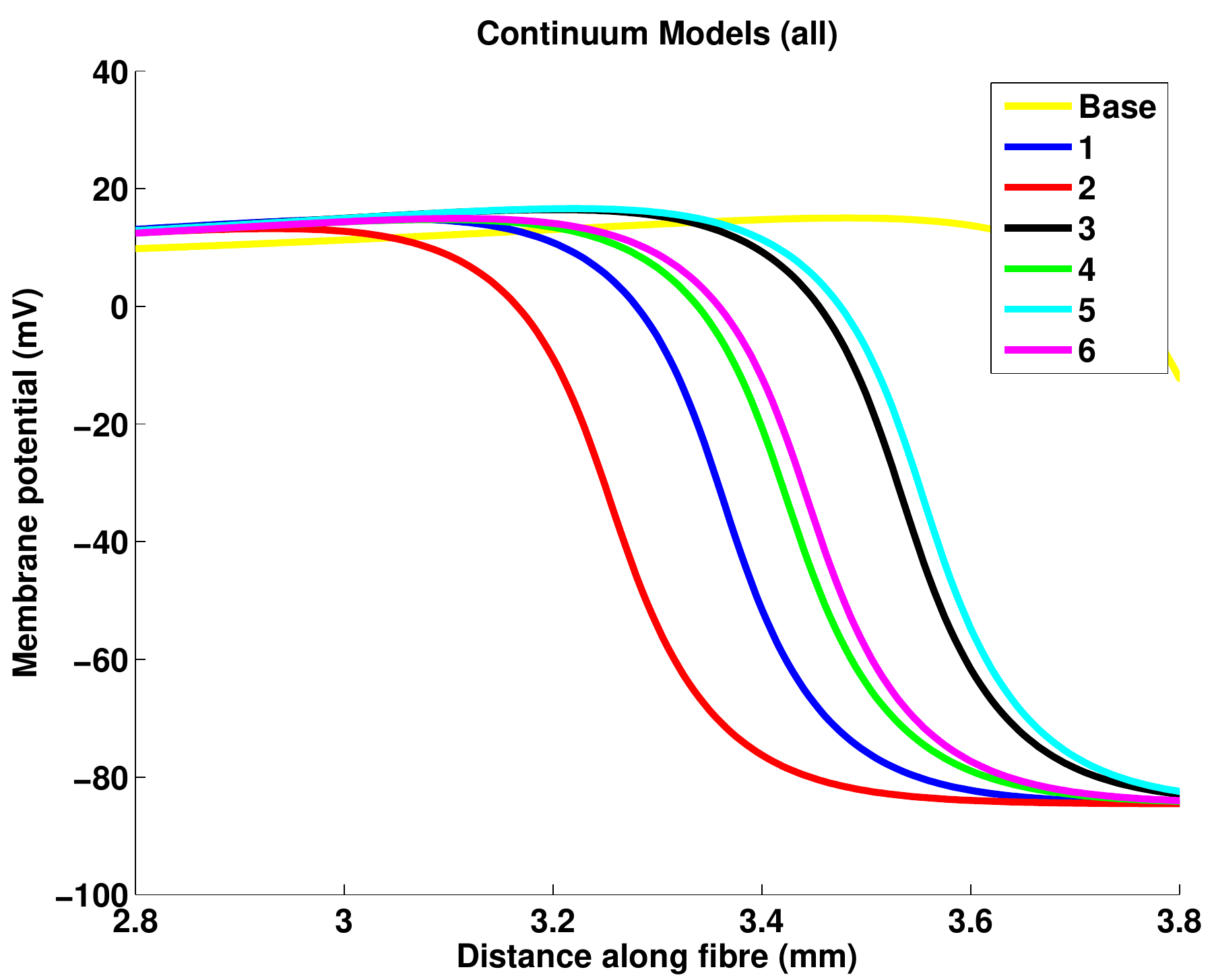}}
\subfloat[]{\label{DiscreteZoom}\includegraphics[width = 0.45\linewidth,height =
0.3\linewidth]{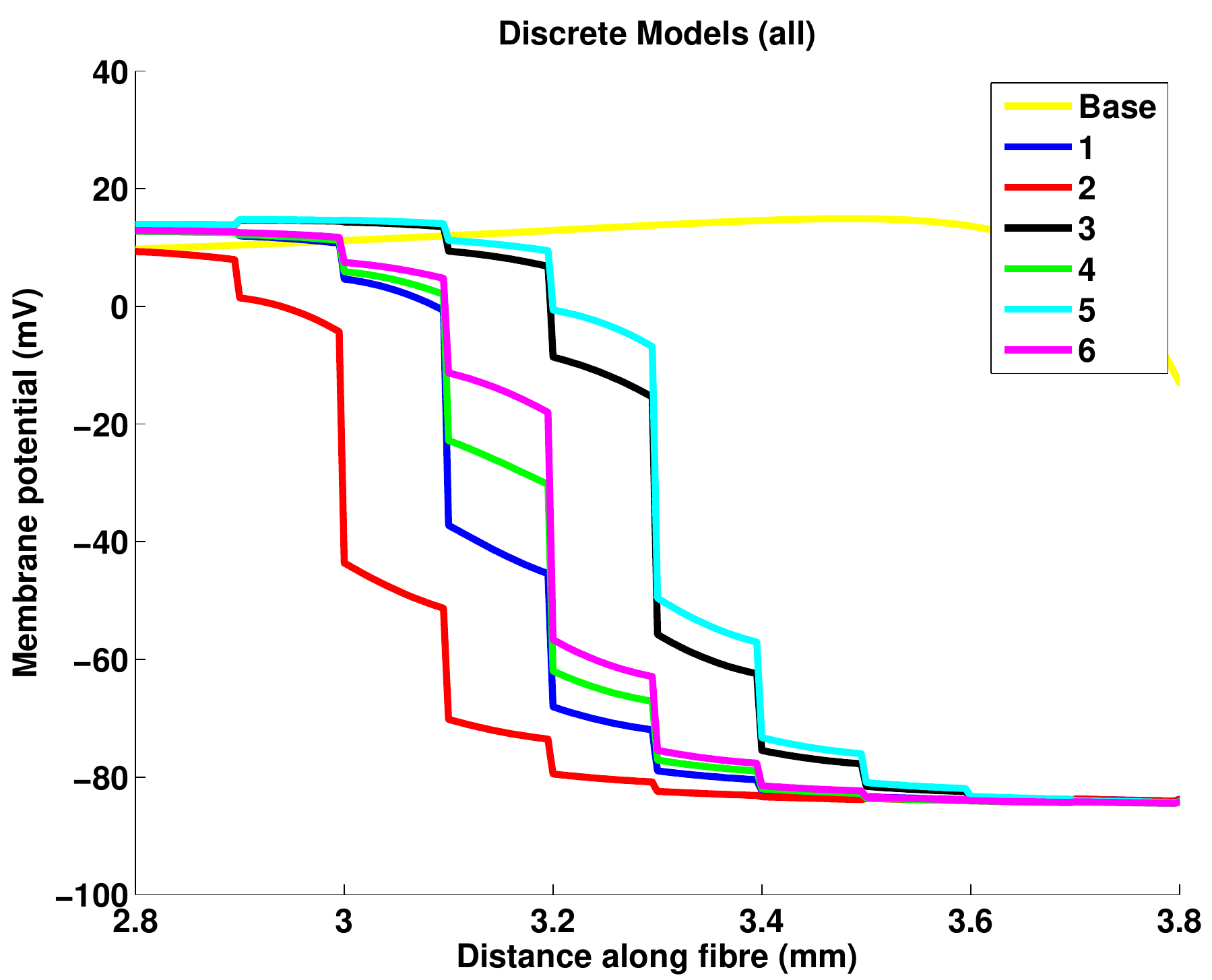}}
\caption{\emph{A magnification of the plots shown in Figure \ref{CompareAll}.}}
\label{CompareAllZoom}
\end{center}
\end{figure}

In Figure \ref{IntroduceGapJunctions} we take spatial snapshots of the results of our simulations at two separate time points --- 15 ms, seen towards the left of each subplot, and 30 ms, seen at the right of each subplot --- for two of the models mentioned above, in both the discrete and continuum formulations of the problem. We compare our Base model in which gap junctions are not modelled, to Model 1 in which the gap junctions are modelled as a region of reduced conductivity whose membrane properties are identical to those of the remainder of the cell.

As expected, in the absence of gap junctions the discrete and continuum models give near-identical solutions, as seen in Figure \ref{Base}. In such a situation the problems highlighted in the introduction concerning the derivation of the continuum model are not applicable, and thus the continuum system provides an accurate representation of the discrete problem. However, when gap junctions are introduced it is seen in Figure \ref{Model1} that the propagation speed in the continuum model does not match that of the discrete model. The conduction velocity of the wave for the discrete problem is noticeably smaller than that for the continuum problem, as observed by the small discrepancy in the solutions after 15ms and the larger discrepancy after 30ms.

This discrepancy occurs because of the rapid spatial variation in the membrane potential in the discrete case as the action potential propagates through a gap junction. Here, the solution lengthscale is of the same order as an individual cell, and thus our key assumption when deriving the continuum model, that we may ignore effects at cell-level and below, is no longer true. It is therefore the case that the bidomain equations, when derived using the inherent cell-level parameters of our system, cannot be used as an accurate representation of the propagation of the action potential if we include the effects of gap junctions in our discrete model.

In addition, we also notice that the discrete simulations in the presence of gap junctions display the form of `stepped' action potential that is seen experimentally in \cite{rohr_2004}. As this is not seen in the absence of gap junctions, we conclude that there should be some representation of a gap junction structure in a model of cardiac electrophysiology in order to capture this more detailed form of the propagated action potential. It is worth noting that the continuum model, in fact \emph{any} continuum model, is unable to replicate such behaviour --- by its nature it cannot have quantities, in this case the intracellular conductivity, that vary on the level of single cells.

\subsection{Comparing implementations of gap junctions}

Having seen that gap junctions change the results of both discrete and continuum simulations of cardiac electrophysiology, ultimately causing the solutions of the two types of model to diverge, we wish to see if the precise nature of the implementation of gap junctions further affects the characteristics of solutions. To that end, in Figure \ref{CompareAll} we plot the results of simulations of both continuum (left-hand figure) and discrete (right-hand figure) versions of each model specified in Table \ref{Models} taken at a time of 30 ms. A magnification of Figure \ref{CompareAll} is given in Figure \ref{CompareAllZoom}.

Whilst we see a difference in the position of the propagating wave --- and thus the underlying wavespeed --- between each of the implementations of gap junctions, this change is much smaller than the initial change brought by the introduction of gap junctions over our Base model.  This suggests that the major cause of the discrepancy between continuum and discrete solutions is the sharp change in conductivity that we have between the cell and the gap junction.

Considering the plots in more detail, we see in Figure \ref{CompareAllZoom} that switching the ionic current off on the gap junction membrane --- going from Model 1 to 2, Model 3 to 4 or Model 5 to 6 --- slows down the propagated wave as expected, though by an equal amount in the continuum and discrete cases. Reducing the capacitance of the gap junction membrane --- Model 1 to 3 and Model 2 to 4 --- again slows down propagation, this time by a larger amount. However, further reducing the capacitance to zero --- Model 3 to 5 and Model 4 to 6 --- has a negligible effect on solutions. More importantly, we can see that such changes in the results of the discrete model are mirrored in the continuum formulation of the problem, specifically the associated change in propagation speed.

It is also clear from Figure \ref{CompareAllZoom} how the steepness of the propagating wave varies from continuum to discrete models --- in the continuum case the wave is moderately steep for the entirety of the upstroke, whereas in the discrete case the wave is fairly shallow as it passes through each cell, and extremely steep inside the gap junction. This reinforces our statement that gap junctions cause rapid spatial variation in the potential.


\section{Conclusion}

The implementation of gap junctions into a standard model of cardiac electrophysiology causes a discrepancy to occur between results of simulations of discrete and continuum versions of the system. This is due to the rapid spatial variation in the membrane potential inside a gap junction that is caused by the concomitant large decrease in conductivity in such a region. Given this, it is not possible to model the contribution and effect of gap junctions on cardiac electrophysiology using a continuum system, and we suggest that a hybrid method --- using the discrete model around the upstroke of the propagated wave, and a continuum model elsewhere --- may enable us to retain accuracy and characterisation of solutions whilst increasing computational tractability.

With regard to the precise implementation of gap junctions, we have seen that solutions will depend on the value of the capacitance of the gap junction membrane, with an adapted version of the continuum system matching the changes predicted by the discrete system. These results suggest it is important to have an accurate value for the capacitance of the gap junction membrane when conducting simulations of cardiac electrophysiology.

\section*{Acknowledgements} 

We would like to thank the referees for their insight and helpful comments, especially with regard to recent papers in the field of hybrid cardiac modelling.

Doug Bruce is supported by an EPSRC grant to the Life Sciences Interface Doctoral Training Centre.


\bibliographystyle{eptcs}
\bibliography{GapJunctions}

\end{document}

%% file: fulldomain.latex
\setlength{\unitlength}{1184sp}%
\begingroup\makeatletter\ifx\SetFigFont\undefined%
\gdef\SetFigFont#1#2#3#4#5{%
  \reset@font\fontsize{#1}{#2pt}%
  \fontfamily{#3}\fontseries{#4}\fontshape{#5}%
  \selectfont}%
\fi\endgroup%
\begin{picture}(10824,8583)(589,-8332)
\thinlines
{\color[rgb]{0,0,0}\put(1201,-1561){\line( 1, 0){9600}}
}%
{\color[rgb]{0,0,0}\put(1201,-3961){\line( 1, 0){9600}}
}%
{\color[rgb]{0,0,0}\put(1201,-5161){\line( 1, 0){9600}}
}%
{\color[rgb]{0,0,0}\put(1201,-2761){\line( 1, 0){9600}}
}%
{\color[rgb]{0,0,0}\put(3601,-1561){\line( 0,-1){1200}}
}%
{\color[rgb]{0,0,0}\put(6001,-1561){\line( 0,-1){1200}}
}%
{\color[rgb]{0,0,0}\put(8401,-1561){\line( 0,-1){1200}}
}%
{\color[rgb]{0,0,0}\put(3601,-3961){\line( 0,-1){1200}}
}%
{\color[rgb]{0,0,0}\put(6001,-3961){\line( 0,-1){1200}}
}%
{\color[rgb]{0,0,0}\put(8401,-3961){\line( 0,-1){1200}}
}%
{\color[rgb]{0,0,0}\put(10801,-361){\line(-1, 0){9600}}
}%
{\color[rgb]{0,0,0}\put(1201,-6361){\line( 1, 0){9600}}
}%
{\color[rgb]{0,0,0}\put(1201,-1561){\line( 0,-1){ 75}}
\put(1201,-1636){\line( 0,-1){ 75}}
\put(1201,-1711){\line( 0,-1){ 75}}
\put(1201,-1786){\line( 0,-1){ 75}}
\put(1201,-1861){\line( 0,-1){ 75}}
\put(1201,-1936){\line( 0,-1){ 75}}
\put(1201,-2011){\line( 0,-1){ 75}}
\put(1201,-2086){\line( 0,-1){ 75}}
\put(1201,-2161){\line( 0,-1){ 75}}
\put(1201,-2236){\line( 0,-1){ 75}}
\put(1201,-2311){\line( 0,-1){ 75}}
\put(1201,-2386){\line( 0,-1){ 75}}
\put(1201,-2461){\line( 0,-1){ 75}}
\put(1201,-2536){\line( 0,-1){ 75}}
\put(1201,-2611){\line( 0,-1){ 75}}
\put(1201,-2686){\line( 0,-1){ 75}}
}%
{\color[rgb]{0,0,0}\put(10801,-3961){\line( 0,-1){1200}}
}%
{\color[rgb]{0,0,0}\put(1201,-5161){\line( 0, 1){1200}}
}%
{\color[rgb]{0,0,0}\put(1201,-1561){\line(-1, 0){ 75}}
\put(1126,-1561){\line(-1, 0){ 75}}
\put(1051,-1561){\line(-1, 0){ 75}}
\put(976,-1561){\line(-1, 0){ 75}}
\put(901,-1561){\line(-1, 0){ 75}}
\put(826,-1561){\line(-1, 0){ 75}}
\put(751,-1561){\line(-1, 0){ 75}}
\put(676,-1561){\line(-1, 0){ 75}}
}%
{\color[rgb]{0,0,0}\put(1201,-2761){\line(-1, 0){600}}
}%
{\color[rgb]{0,0,0}\put(1201,-3961){\line(-1, 0){600}}
}%
{\color[rgb]{0,0,0}\put(1201,-5161){\line(-1, 0){600}}
}%
{\color[rgb]{0,0,0}\put(1201,-6361){\line(-1, 0){600}}
}%
{\color[rgb]{0,0,0}\put(1201,-361){\line(-1, 0){600}}
}%
{\color[rgb]{0,0,0}\put(10801,-1561){\line( 0,-1){1200}}
}%
{\color[rgb]{0,0,0}\put(10801,-361){\line( 1, 0){ 75}}
\put(10876,-361){\line( 1, 0){ 75}}
\put(10951,-361){\line( 1, 0){ 75}}
\put(11026,-361){\line( 1, 0){ 75}}
\put(11101,-361){\line( 1, 0){ 75}}
\put(11176,-361){\line( 1, 0){ 75}}
\put(11251,-361){\line( 1, 0){ 75}}
\put(11326,-361){\line( 1, 0){ 75}}
}%
{\color[rgb]{0,0,0}\put(10801,-1561){\line( 1, 0){600}}
}%
{\color[rgb]{0,0,0}\put(10801,-2761){\line( 1, 0){600}}
}%
{\color[rgb]{0,0,0}\put(10801,-3961){\line( 1, 0){600}}
}%
{\color[rgb]{0,0,0}\put(10801,-5161){\line( 1, 0){600}}
}%
{\color[rgb]{0,0,0}\put(10801,-6361){\line( 1, 0){525}}
}%
{\color[rgb]{0,0,0}\put(11326,-6361){\line( 1, 0){ 75}}
}%
{\color[rgb]{0,0,0}\put(1201,-361){\line( 0, 1){600}}
}%
{\color[rgb]{0,0,0}\put(3601,-361){\line( 0, 1){600}}
}%
{\color[rgb]{0,0,0}\put(6001,-361){\line( 0, 1){525}}
}%
{\color[rgb]{0,0,0}\put(8401,-361){\line( 0, 1){600}}
}%
{\color[rgb]{0,0,0}\put(6001,164){\line( 0, 1){ 75}}
}%
{\color[rgb]{0,0,0}\put(10801,-361){\line( 0, 1){600}}
}%
{\color[rgb]{0,0,0}\put(1201,-6361){\line( 0,-1){600}}
}%
{\color[rgb]{0,0,0}\put(3601,-6361){\line( 0,-1){600}}
}%
{\color[rgb]{0,0,0}\put(6001,-6361){\line( 0,-1){600}}
}%
{\color[rgb]{0,0,0}\put(8401,-6361){\line( 0,-1){675}}
}%
{\color[rgb]{0,0,0}\put(10801,-6361){\line( 0,-1){600}}
}%
\thicklines
{\color[rgb]{1,0,0}\put(3601,-3361){\framebox(2400,2400){}}
}%
{\color[rgb]{0,0,0}\put(9601,-8161){\vector( 0, 1){1200}}
}%
{\color[rgb]{0,0,0}\put(7501,-3286){\vector(-1, 0){1350}}
}%
{\color[rgb]{0,0,0}\put(9601,-8161){\vector( 1, 0){1200}}
}%
\put(10876,-8236){\makebox(0,0)[lb]{\smash{{\SetFigFont{6}{7.2}{\rmdefault}{\bfdefault}{\updefault}{\color[rgb]{0,0,0}$x$}%
}}}}
\put(9076,-2161){\makebox(0,0)[lb]{\smash{{\SetFigFont{8}{9.6}{\rmdefault}{\bfdefault}{\updefault}{\color[rgb]{0,0,0}$\sigma_i$, $\phi_i$}%
}}}}
\put(9451,-6811){\makebox(0,0)[lb]{\smash{{\SetFigFont{6}{7.2}{\rmdefault}{\bfdefault}{\updefault}{\color[rgb]{0,0,0}$y$}%
}}}}
\put(9076,-1036){\makebox(0,0)[lb]{\smash{{\SetFigFont{8}{9.6}{\rmdefault}{\bfdefault}{\updefault}{\color[rgb]{0,0,0}$\sigma_e$, $\phi_e$}%
}}}}
\put(7576,-3361){\makebox(0,0)[lb]{\smash{{\SetFigFont{8}{9.6}{\rmdefault}{\bfdefault}{\updefault}{\color[rgb]{0,0,0}$\Omega$}%
}}}}
\end{picture}%

%% file: subunit.latex
%
%
\setlength{\unitlength}{2368sp}%
\begingroup\makeatletter\ifx\SetFigFont\undefined%
\gdef\SetFigFont#1#2#3#4#5{%
  \reset@font\fontsize{#1}{#2pt}%
  \fontfamily{#3}\fontseries{#4}\fontshape{#5}%
  \selectfont}%
\fi\endgroup%
\begin{picture}(5655,4377)(1261,-4876)
\thinlines
{\color[rgb]{0,0,0}\put(2401,-1711){\line( 1, 0){3600}}
}%
{\color[rgb]{0,0,0}\put(2401,-2911){\line( 1, 0){3600}}
}%
{\color[rgb]{0,0,0}\put(2401,-4111){\framebox(3600,3600){}}
}%
{\color[rgb]{0,0,0}\put(2401,-4561){\vector(-1, 0){  0}}
\put(2401,-4561){\vector( 1, 0){3600}}
}%
{\color[rgb]{0,0,0}\put(1651,-4111){\vector( 0,-1){  0}}
\put(1651,-4111){\vector( 0, 1){3600}}
}%
{\color[rgb]{0,0,0}\put(6826,-2311){\vector(-1, 0){1425}}
}%
{\color[rgb]{0,0,0}\put(2251,-2986){\vector( 0,-1){  0}}
\put(2251,-2986){\vector( 0, 1){1275}}
}%
{\color[rgb]{0,0,0}\put(6826,-3511){\vector(-1, 0){1425}}
}%
{\color[rgb]{0,0,0}\put(4951,-1711){\vector( 0, 1){600}}
}%
\put(6901,-2386){\makebox(0,0)[lb]{\smash{{\SetFigFont{12}{14.4}{\familydefault}{\mddefault}{\updefault}{\color[rgb]{0,0,0}$\Omega_i$}%
}}}}
\put(4126,-4861){\makebox(0,0)[lb]{\smash{{\SetFigFont{12}{14.4}{\rmdefault}{\mddefault}{\updefault}{\color[rgb]{0,0,0}L}%
}}}}
\put(6901,-3586){\makebox(0,0)[lb]{\smash{{\SetFigFont{12}{14.4}{\familydefault}{\mddefault}{\updefault}{\color[rgb]{0,0,0}$\Omega_e$}%
}}}}
\put(1276,-2461){\makebox(0,0)[lb]{\smash{{\SetFigFont{12}{14.4}{\familydefault}{\mddefault}{\updefault}{\color[rgb]{0,0,0}$h$}%
}}}}
\put(1876,-2461){\makebox(0,0)[lb]{\smash{{\SetFigFont{12}{14.4}{\familydefault}{\mddefault}{\updefault}{\color[rgb]{0,0,0}$h_1$}%
}}}}
\put(3826,-2761){\makebox(0,0)[lb]{\smash{{\SetFigFont{12}{14.4}{\familydefault}{\mddefault}{\updefault}{\color[rgb]{0,0,0}$\partial \Omega_m$}%
}}}}
\put(3826,-1561){\makebox(0,0)[lb]{\smash{{\SetFigFont{12}{14.4}{\familydefault}{\mddefault}{\updefault}{\color[rgb]{0,0,0}$\partial \Omega_m$}%
}}}}
\put(4876,-1036){\makebox(0,0)[lb]{\smash{{\SetFigFont{12}{14.4}{\familydefault}{\mddefault}{\updefault}{\color[rgb]{0,0,0}$\textbf{n}$}%
}}}}
\end{picture}%

%% file: gapjcts.latex
%
%
\setlength{\unitlength}{2368sp}%
\begingroup\makeatletter\ifx\SetFigFont\undefined%
\gdef\SetFigFont#1#2#3#4#5{%
  \reset@font\fontsize{#1}{#2pt}%
  \fontfamily{#3}\fontseries{#4}\fontshape{#5}%
  \selectfont}%
\fi\endgroup%
\begin{picture}(5505,3477)(1411,-4426)
\thinlines
{\color[rgb]{0,0,0}\put(6826,-2311){\vector(-1, 0){1425}}
}%
{\color[rgb]{0,0,0}\put(2401,-3661){\framebox(3600,2700){}}
}%
{\color[rgb]{0,0,0}\put(2401,-4036){\vector(-1, 0){  0}}
\put(2401,-4036){\vector( 1, 0){3600}}
}%
{\color[rgb]{0,0,0}\put(1801,-3661){\vector( 0,-1){  0}}
\put(1801,-3661){\vector( 0, 1){2700}}
}%
{\color[rgb]{0,0,0}\put(6826,-3211){\vector(-1, 0){1425}}
}%
{\color[rgb]{0,0,0}\put(2401,-2986){\vector(-1, 0){  0}}
\put(2401,-2986){\vector( 1, 0){300}}
}%
{\color[rgb]{0,0,0}\put(3976,-2311){\vector(-1, 0){1425}}
}%
{\color[rgb]{0,0,0}\put(2701,-1861){\line( 0,-1){900}}
}%
{\color[rgb]{0,0,0}\put(2401,-1861){\line( 1, 0){3600}}
}%
{\color[rgb]{0,0,0}\put(2251,-2761){\vector( 0,-1){  0}}
\put(2251,-2761){\vector( 0, 1){900}}
}%
{\color[rgb]{0,0,0}\put(2401,-2761){\line( 1, 0){3600}}
}%
\put(1876,-2461){\makebox(0,0)[lb]{\smash{{\SetFigFont{12}{14.4}{\familydefault}{\mddefault}{\updefault}{\color[rgb]{0,0,0}$h_1$}%
}}}}
\put(6901,-2386){\makebox(0,0)[lb]{\smash{{\SetFigFont{12}{14.4}{\familydefault}{\mddefault}{\updefault}{\color[rgb]{0,0,0}$\sigma_i$}%
}}}}
\put(2554,-3361){\makebox(0,0)[lb]{\smash{{\SetFigFont{12}{14.4}{\familydefault}{\mddefault}{\updefault}{\color[rgb]{0,0,0}$\delta$}%
}}}}
\put(4126,-4411){\makebox(0,0)[lb]{\smash{{\SetFigFont{12}{14.4}{\rmdefault}{\mddefault}{\updefault}{\color[rgb]{0,0,0}L}%
}}}}
\put(4126,-2386){\makebox(0,0)[lb]{\smash{{\SetFigFont{12}{14.4}{\familydefault}{\mddefault}{\updefault}{\color[rgb]{0,0,0}$\sigma_g$}%
}}}}
\put(6901,-3361){\makebox(0,0)[lb]{\smash{{\SetFigFont{12}{14.4}{\familydefault}{\mddefault}{\updefault}{\color[rgb]{0,0,0}$\sigma_e$}%
}}}}
\put(1426,-2461){\makebox(0,0)[lb]{\smash{{\SetFigFont{12}{14.4}{\familydefault}{\mddefault}{\updefault}{\color[rgb]{0,0,0}$h$}%
}}}}
\end{picture}%